# Modélisation des facteurs influençant la performance de la chaîne logistique

Omar Sakka[1], Valerie Botta-Genoulaz[1], Lorraine Trilling[1]

[1] Universite de Lyon, INSA-Lyon, LIESP
19, avenue Jean Capelle, dpt GI, Bat Jules Verne, F-69621 Villeurbanne, France

{omar.sakka, valerie.botta, lorraine.trilling}@insa-lyon.fr

*Résumé* - Les entreprises cherchent de plus en plus à améliorer leur performance industrielle en termes de coût, de délais, d'adaptabilité, de variété et de traçabilité. A ce besoin correspond la nécessité pour les entreprises de collaborer et de renforcer leurs dispositifs de coordination. L'échange des informations devient alors une question stratégique : Quelle est la nature des informations susceptibles d'être partagées avec ses clients et ses fournisseurs ? Quel est l'impact sur la performance de l'entreprise et sur l'ensemble de la chaîne ? Il est indispensable pour l'entreprise d'identifier les informations dont l'échange contribue à sa performance et de maîtriser ses flux d'informations. En se basant sur une analyse de la littérature scientifique récente, ce travail vise à identifier les principales tendances de pratiques d'échanges d'informations et de collaboration aboutissant à la performance et propose un modèle de facteurs relationnels et technologiques contribuant à la performance des chaines logistiques.

*Abstract* - Improvement of industrial performance such as cost, lead-time, adaptability, variety and traceability is the major finality of companies. At this need corresponds the necessity to collaborate and to strengthen their coordination mechanisms. Information exchange becomes then a strategic question: what is the nature of the information that can be shared with customers and suppliers? Which impact on the performance of a company is expectable? What about the performance of the whole supply chain? It is essential for a company to identify the information whose exchange contributes to its performance and to control its information flows. This study aims to release from the literature the main tendencies of collaboration practices and information exchanges leading to the performance and to propose a model of hypothesis gathering these practices.

*Mots clés* - chaîne logistique, performance, pratiques, collaboration, échange d'information.

## 1 Introduction

Pour faire face aux fortes pressions de la concurrence mondiale, les entreprises sont constamment à la recherche de nouvelles façons d'améliorer la performance de leur chaîne logistique afin de réduire les coûts, d'améliorer la qualité et d'augmenter la productivité. Il est devenu évident, pour un acteur industriel, que c'est dans le cadre de l'atteinte d'un objectif global et la considération de l'ensemble de la chaîne logistique, du fournisseur du fournisseur au client du client, que se trouve la clé de réussite. Passer d'une activité centrée sur le produit à une activité qui doit répondre en permanence aux besoins des autres membres de la chaîne n'est pas une mutation aisée. Les frontières d'une entreprise sont alors dépassées si bien que les autres partenaires sont considérés comme des membres d'une structure étendue. Pourtant se mettre brusquement à échanger, partager de l'information, des compétences avec d'autres acteurs de la chaîne n'est pas chose facile. A cette mutation organisationnelle correspond la nécessité pour les entreprises de renforcer significativement leur dispositif de coordination et de collaboration. Le partage et l'échange des informations deviennent alors des questions stratégiques.

C'est dans le contexte des pratiques d'échange d'information et de collaboration que nous intervenons.

Dans cette étude nous nous sommes intéressés à identifier, dans un premier temps, les principaux facteurs et pratiques d'échanges d'informations et de collaboration qui peuvent contribuer à la performance de la chaîne logistique en général et la performance des acteurs de la chaîne en particulier et dans un deuxième temps, à construire un modèle générique qui se base sur ces différents facteurs. Pour ce faire, nous avons été amenés à réaliser une étude bibliographique portant sur les pratiques d'échanges d'informations et les pratiques collaboratives qui aboutissent à la performance. A partir de cette étude nous avons pu modéliser l'ensemble des hypothèses identifiées à partir de la littérature.

La lecture de cet article se fait comme suit : la deuxième section est consacrée à l'étude bibliographique où nous analysons les principaux travaux de recherche récents traitant du sujet. Une synthèse et une classification des facteurs contribuant à la performance des chaînes logistiques sont proposées en section trois. Nous développons dans la section quatre un modèle générique regroupant les facteurs et les pratiques d'échanges d'informations et de collaboration qui contribuent à la performance. Nous terminons cet article par la discussion de notre travail et nous concluons par les points forts à retenir de notre contribution ainsi que les perspectives liées à ce travail.

## 2 ETAT DE L'ART

Dans un contexte où les entreprises cherchent à améliorer leur performance industrielle en termes de coût, de délais, d'adaptabilité, de variété et de traçabilité, le degré de collaboration et d'échange d'informations entre partenaires devient un élément essentiel à considérer au sein de toute chaîne logistique [Campagne, 2006].

De nombreux travaux cherchent à améliorer la performance des chaînes logistiques. Nous nous sommes basés dans nos recherches sur les principaux travaux réalisés ces dernières années pour faire ressortir les facteurs contribuant à la performance des chaînes logistiques. La plupart de ces travaux se basent sur des enquêtes réalisées auprès d'entreprises. Les auteurs tentent à travers leurs modèles et hypothèses de renforcer les dispositifs de collaboration et de coopération pour contribuer à la performance de l'entreprise en particulier et à la performance de la chaîne logistique en général. La validation des hypothèses par les auteurs se base soit sur l'application de modèles mathématiques et statistiques, comme les modèles d'équations structurelles (par exemple avec l'outil RAMONA [Paulraj et al., 2008]), soit sur des simulations (par exemple avec Arena [Chen et al., 2007]). Ainsi on trouve différentes stratégies et contributions qui se basent en général sur les technologies d'information, le partage et l'échange d'information, les pratiques de la chaîne logistique, la communication inter-organisationnelle et la collaboration comme des facteurs essentiels pour aboutir à la performance des entreprises.

Paulraj et al. [Paulraj et al., 2008] ont considéré la communication inter-organisationnelle comme un facteur critique pour la collaboration stratégique entre entreprises. Ils définissent la communication inter-organisationnelle comme une compétence relationnelle qui peut fournir un avantage stratégique pour les partenaires de la chaîne et améliorer ainsi leur performance.

Focalisés sur la collaboration et l'utilisation de la technologie d'information, Chen et al. [Chen et al., 2007] se sont basés sur le CPFR (Collaborative Planning Forecasting and Replenishment), une collection de pratiques qui vise à réduire radicalement les stocks et les dépenses et tente aussi d'augmenter le service client, pour simuler quatre scénarios de collaboration entre un détaillant et un fournisseur. Les auteurs ont pu montrer que l'utilisation de l'IT ne peut pas elle seule assurer les gains pour les partenaires de la chaîne et que la mutualisation est un facteur essentiel pour atteindre ce but.

En se basant toujours sur l'utilisation de la technologie d'information et dans le but d'améliorer d'autres types de performance- la performance du marché et la performance financière- Seggie et al. [Seggie et al., 2006] ont exploré deux aspects des technologies d'information : l'intégration des systèmes d'information inter-organisationnels et l'alignement de la technologie d'information entre les membres de la chaîne (l'alignement de la technologie d'information est le degré de la compatibilité de la technologie d'information de l'entreprise avec celles des autres partenaires de la chaîne). Ces deux aspects d'IT engendrent l'amélioration du capital de marque de l'entreprise qui, à son tour, influe positivement sur la performance du marché en termes de développement, de partage et d'augmentation de ventes. Ceci engendre directement l'amélioration de la performance financière de l'entreprise.

Plus généralement et en cherchant toujours à améliorer la performance de l'entreprise, Sanders [Sanders, 2007] proposent un modèle de relation entre l'usage des technologies de e-business, la collaboration et la performance organisationnelle. En effet, la croissance rapide des technologies d'information, et surtout d'Internet et du web, a permis la collaboration et l'intégration en temps réel entre les partenaires de la chaîne. Cette étude a confirmé l'impact positif de l'usage des technologies de e-business sur la performance de l'entreprise et sur la collaboration intra et inter-organisationnelle. Cette dernière influe à son tour directement et fortement sur la performance. Cette contribution met en valeur l'importance pour les entreprises de promouvoir la collaboration en interne et d'investir dans les techniques de e-business.

De nouvelles notions sont apparues avec [Yang et al., 2008]. Les auteurs se sont intéressés à la stabilité relationnelle dans les alliances de la chaîne logistique, l'engagement relationnel et la confiance envers les fournisseurs. Dans leur travail, les auteurs abordent aussi la notion de l'alliance coopérative entre les fournisseurs et les clients. Cette alliance permet aux entreprises de partager les risques financiers, d'améliorer la qualité de service, d'augmenter la productivité et de réduire les coûts. A travers cette étude, les auteurs ont pu montrer l'importance, aux yeux des entreprises, de l'engagement relationnel et de la confiance envers les fournisseurs pour développer une relation stable dans le cadre de l'alliance de la chaîne logistique et ainsi aboutir à la performance.

D'autres types de performance sont abordés dans la littérature tels que la performance de la livraison, la performance du fournisseur, la performance du client, etc. Dans le travail de Zhou et Benton [Zhou et Benton, 2007] qui se base essentiellement sur trois aspects de partage d'information (la technologie de support de partage d'information, le contenu de l'information et la qualité de l'information) et les pratiques de la chaîne logistique (pratiques de livraison, pratiques de production et de planification), les auteurs se sont intéressés à la performance de la chaîne logistique et particulièrement à la performance de la livraison caractérisée par la livraison à temps, le taux d'accomplissement de la commande et la fiabilité de la livraison. Ce travail a montré que le partage d'information, avec ses trois aspects, et le dynamisme de la chaîne logistique (vitesse de changement dans les produits et les processus à la fois) influent fortement sur l'efficacité des pratiques de la chaîne.

D'autres travaux se sont intéressés à tester l'influence d'autres types de pratiques sur la performance de la chaîne logistique. Li et al. [Li et al., 2006] ont cité comme pratiques pour la gestion de la chaîne : le choix stratégique de fournisseurs, la fondation d'une relation à long terme avec le client, le degré de partage d'information, la qualité de partage d'information et la différentiation retardée. Ces pratiques influent positivement sur la performance financière et la performance du marché et créent aussi un avantage compétitif pour l'entreprise en termes de prix, qualité, fiabilité de la livraison, innovation de produit et délai de mise en marché. Chowa et al. [Chowa et al., 2008] ajoutent d'autres pratiques comme la communication, l'intégration et la gestion des services du client, les caractéristiques de la chaîne logistique, le partage d'information et précisent que même si ces pratiques diffèrent selon la zone géographique, elles influent toujours positivement sur la performance globale de la chaîne [Chowa et al., 2008].

Parmi toutes les pratiques étudiées dans la littérature, l'échange et le partage d'information restent les pratiques les plus présentes. De nombreux auteurs tels que [Gaonkar et Viswanadham, 2001], [Laux et al., 2004], [Liu et Kumar, 2003], [Sahin et Robinson, 2005], [Shore, 2001], [Llerena et al., 2006], [Gruat-La-Forme, 2007] et [Botta-Genoulaz et al., 2005], ont cherché à identifier les informations susceptibles d'être partagées, d'évaluer l'impact de ces échanges et la répartition du gain attendu ou encore de discuter des questions techniques et informatiques soulevées par ce partage. On peut ainsi dégager six types d'informations à échanger dans la chaîne logistique : des informations sur les produits, sur les ressources, sur les stocks, sur les délais, sur les demandes et sur les données de planification.

En pratique, les dirigeants des entreprises choisissent de mettre en œuvre soit un partage d'information efficace soit d'améliorer les pratiques de la chaîne logistique, compte tenu de la limitation des ressources qui empêche de poursuivre les deux à la fois. Comme solution, Zhou et Benton [Zhou et Benton, 2007] proposent la standardisation des processus de la chaîne logistique. Grâce à cette standardisation, les entreprises pourront aboutir à un partage efficace des informations avec leurs partenaires, augmentant ainsi leur performance en diminuant les incertitudes, ce qui donne un aspect réactif à la chaîne logistique.

On trouve aussi dans la littérature de nouvelles stratégies et pratiques qui visent d'améliorer la performance de l'entreprise. Dans le travail de [Modi et Mabert, 2007], les auteurs cherchent à améliorer la performance du fournisseur considéré comme une ressource critique pour l'entreprise, fournissant directement et indirectement de la matière et du service. En outre la qualité et le coût des produits ou des services offerts sur le marché dépendent non seulement de la capacité de l'entreprise mais aussi du réseau des fournisseurs pour les entreprises. Alors pour rester compétitive, l'entreprise implémente de plus de plus des programmes de développement de ses fournisseurs pour maintenir la performance de la chaîne. Klein [Klein, 2007] traite de la relation entre les fournisseurs de services et les clients pour créer un environnement collaboratif. Cette relation se base essentiellement sur la confiance puisque cette dernière joue un rôle fondamental dans l'évolution des relations inter-organisationnelles dans la chaîne logistique et elle est considérée comme le composant noyau pour des relations persistantes et des alliances stratégiques [Ganesan, 1994], [Gulati, 1995]. Cette relation se base aussi sur l'utilisation des technologies e-business, le niveau de personnalisation des solutions par les clients et l'accès en temps réel aux applications.

Fynes et al. [Fynes et al., 2005] ont étudié les relations entre les acteurs de la chaîne logistique dans le but d'améliorer la satisfaction du client. Les auteurs affirment que des dimensions telles que la confiance, l'adaptation, l'interdépendance, la coopération, la communication et l'engagement accompagnent et renforcent les relations dans la chaîne. Ils définissent ainsi la qualité de relation dans la chaîne logistique comme étant le degré d'engagement de deux parties dans une relation de travail active, à long terme et la mise en service d'indicateurs de communication, de confiance, d'adaptation, d'engagement, d'interdépendance et de coopération.

Le point commun des travaux présentés ci-dessus est qu'ils cherchent à améliorer la performance de la chaîne logistique tout en suivant plusieurs stratégies ou modèles. L'usage des technologies d'information, le partage et l'échange d'informations et de connaissances, la communication inter-organisationnelle, la confiance, la coopération et la collaboration intra- et inter-organisationnelle forment des facteurs indispensables pour la performance de la chaîne. Cependant une entreprise doit être en mesure de définir clairement ses besoins en performance afin d'identifier quelles sont les meilleures pratiques répondant à ces besoins.

Dans la partie suivante, nous allons présenter d'une manière plus précise et exhaustive l'ensemble des facteurs relevés de la littérature que l'on a pu regrouper et qui aboutissent à la performance de la chaîne logistique.

## 3 CLASSIFICATION DES FACTEURS INFLUENÇANT LA PERFORMANCE

D'après nos analyses de la littérature, nous nous apercevons que chaque travail de recherche vise à identifier les influences de quelques pratiques collaboratives ou techniques sur un ou deux types de performance, en se basant presque exclusivement sur la formulation puis le test d'hypothèses. Le but de cette section est de regrouper et de classer l'ensemble des pratiques et des méthodes identifiées dans la littérature comme aboutissant à l'amélioration de la performance.

Nous avons pu extraire cinquante sept hypothèses des articles étudiés [Sakka et al., 2008]. Toutes ces hypothèses ont été testées par les experts du domaine à travers des enquêtes auprès de nombreuses entreprises.

Pour classer ces hypothèses, nous avons cherché dans un premier temps à caractériser la nature de chacune d'elles. Cela nous a amené à exprimer les hypothèses sous la forme « A *a un effet positif sur* B » où A est une caractéristique dite « facteur » et B est une caractéristique dite « impact/résultat », en respectant tant que faire se peut l'idée originale des auteurs. Exemple : « *L'usage des technologies d'information a un effet positif sur une communication inter-organisationnelle* ». Dans cet exemple « l'usage de la technologie d'information » est le facteur et « la communication inter-organisationnelle » est l'impact/résultat. On peut trouver des hypothèses qui ont été vérifiées et d'autres dont les expérimentations des auteurs n'ont pas permis d'aboutir à la vérification. Ainsi, une hypothèse est vérifiée si la caractéristique A peut engendrer ou améliorer la caractéristique B. Si la caractéristique A n'engendre pas ou n'améliore pas forcément la caractéristique B, on dira que l'hypothèse n'est pas vérifiée (à ne pas confondre avec une hypothèse fausse).

Nous avons ensuite regroupé les caractéristiques identifiées en 4 catégories principales : caractéristiques de performance, caractéristiques techniques et choix technologique, caractéristiques relationnelles et contextuelles et caractéristiques de pratiques de la chaîne logistique. Ce qui nous a guidés pour le regroupement des caractéristiques en quatre catégories c'est la nature de la caractéristique interprétée à partir de la contribution originale des auteurs.

### 3.1 *Caractéristiques de performance (10 caractéristiques)*

Cette catégorie regroupe les différentes formes de performance citées dans les travaux étudiés : la performance du client, la performance du fournisseur, la performance financière, la performance du marché, la performance de la livraison, la performance des prestataires de services, la performance d'alliance et la performance organisationnelle. Le but recherché est généralement d'améliorer ces performances. Il arrive donc très souvent que les caractéristiques de performance soient des caractéristiques « impact/résultat » d'une hypothèse testée. A noter que la performance globale de la chaîne logistique est une combinaison de ces différentes performances.

### 3.2 Caractéristiques relationnelles et contextuelles (19 caractéristiques)

Ces caractéristiques forment la plus grande classe puisqu'elles présentent les différentes relations qui peuvent exister entre les acteurs de la chaîne logistique et caractérisent aussi le contexte dans lequel évolue l'entreprise. Ces caractéristiques décrivent ainsi le comportement, la vision, le type de relation et les interactions entre une entreprise et ses partenaires (fournisseurs ou/et clients). Parmi ces caractéristiques nous citons la confiance envers le fournisseur/client, une orientation d'une relation à long terme, la communication inter et intra-organisationnelle, la collaboration…Certaines caractéristiques peuvent être regroupées en sous-classes car elles touchent à des aspects similaires : qualité de relation de la chaîne logistique, confiance, pouvoir.

### 3.3 Caractéristiques techniques et choix technologique (8 caractéristiques)

Cette catégorie se base sur l'utilisation des technologies d'information et regroupe les techniques et les choix technologiques dans l'entreprise : les supports de partage d'information, l'intégration des systèmes interentreprises, la technologie de commerce électronique...

Ce type de caractéristique est indispensable pour l'amélioration de la performance globale de la chaîne logistique vu l'importance des nouvelles technologies de l'information et de la communication.

### 3.4 Caractéristiques de pratiques de la chaîne logistique (9 caractéristiques)

Cette classe de caractéristiques regroupe plusieurs pratiques de la chaîne logistique. Ces pratiques sont définies comme l'ensemble des activités assumées par une organisation pour promouvoir une gestion efficace de sa chaîne logistique [Li et al., 2006]. On trouve ainsi dans cette classe les pratiques de partage d'information, de production, de planification, etc.

### 3.5 Synthèse

Le tableau 1 regroupe les quatre classes de caractéristiques et présente les références bibliographiques étudiées. Chacun des auteurs s'est intéressé à une ou plusieurs caractéristiques. Nous avons mentionné ainsi pour chaque travail de recherche dans combien d'hypothèses une caractéristique apparait, soit en tant que facteur (F) soit en tant que résultat (R). Ce tableau met en évidence certaines interprétations :

- Le but commun de tous ces travaux est d'améliorer un ou plusieurs types de performance.
- Tous les auteurs se sont intéressés d'avantage aux caractéristiques relationnelles qu'aux caractéristiques technologiques/choix technique et aux caractéristiques de pratiques de la chaîne.
- Les caractéristiques techniques et choix technologiques et les caractéristiques de pratiques de la chaîne sont plutôt des caractéristiques facteurs.
- Les caractéristiques relationnelles sont étudiées par tous les auteurs (à l'exception d'un seul) et elles sont plus présentes en facteurs qu'en résultats.

## 4 MODELE DE FACTEURS INFLUENÇANT LA PERFORMANCE

L'identification des hypothèses étudiées dans la littérature et la classification des caractéristiques nous a amené à construire un modèle détaillé des facteurs influençant la performance. Ce modèle (Figure 1) est un graphe comprenant quatre types de sommet (une forme de sommet pour chaque type de caractéristique) et deux types de flèches (une flèche continue représente une hypothèse testée et vérifiée et une flèche interrompue représente une hypothèse testée et non vérifiée).

- Les sommets peuvent avoir l'une des formes suivantes : un rectangle, un rectangle à coins arrondis, un hexagone et un octogone pour présenter respectivement les caractéristiques de pratiques de la chaîne, les caractéristiques techniques/technologiques, les caractéristiques relationnelles et contextuelles et les caractéristiques de performance.
- Les flèches sont de couleurs différentes selon le type de sommets qu'elles relient. Chaque couleur indique l'interaction d'une classe avec une autre.

Chaque classe regroupe des caractéristiques qui peuvent être à la fois « facteurs » (origine d'une flèche) et/ou « résultats » (destination d'une flèche). Les différentes relations entres ces éléments forment les hypothèses. Nous pouvons donc définir une hypothèse comme étant la relation entre deux caractéristiques de même nature ou de nature différente.

**Tableau 1. Synthèse de références bibliographique**

| Références | Caractéristiques |||||||||
|---|---|---|---|---|---|---|---|---|
| | Techniques et choix technologique || Pratiques de la chaîne || Relationnelles || Performance ||
| | F | R | F | R | F | R | F | R |
| [Paulraj et al., 2008] | 2 | | | | 6 | 6 | | 2 |
| [Sanders, 2007] | 3 | | | | 2 | 3 | | 2 |
| [Yang et al., 2008] | | | | | 4 | 3 | | 1 |
| [Klein, 2007] | 5 | 3 | | | 4 | 2 | | 4 |
| [Seggie et al., 2006] | 5 | 5 | | | 4 | 2 | 1 | 3 |
| [Modi et Mabert, 2007] | | | 1 | | 4 | 4 | | 1 |
| [Zhou et Benton, 2007] | 1 | | 7 | 2 | 2 | 1 | | 7 |
| [Li et al., 2006] | | | 5 | | 5 | 5 | | 2 |
| [Fynes et al., 2005] | | | | | 6 | | 2 | 3 |
| [Chowa et al., 2008] | | | 5 | | | | | 1 |
| **Sous-Total** | **16** | **8** | **18** | **2** | **37** | **26** | **3** | **26** |
| **Total** | **24** || **20** || **63** || **29** ||

F : Facteur, R : Résultat

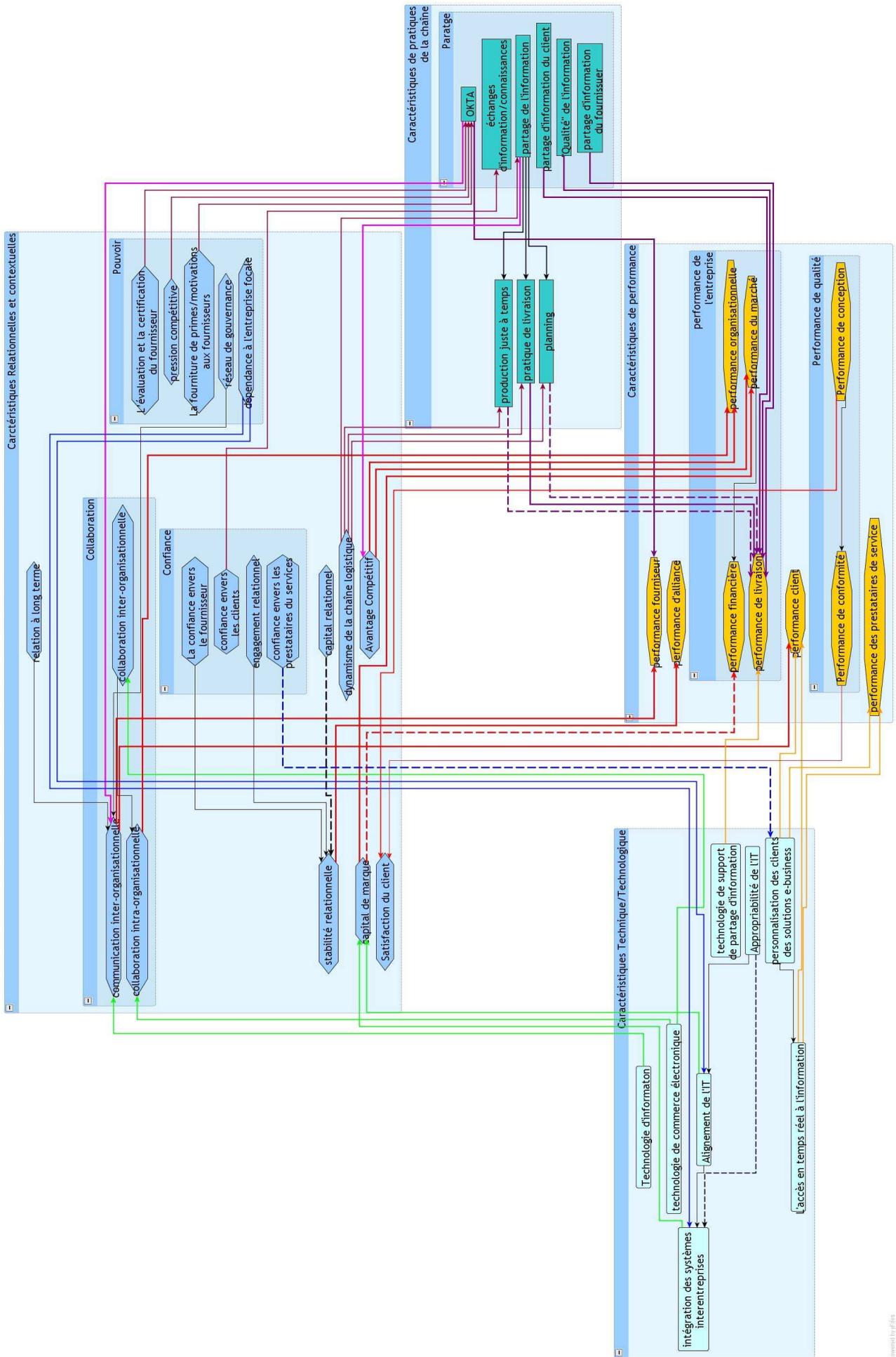

**Figure 1. Modèle détaillé des facteurs influençant la performance**

De plus, une classe de caractéristiques peut regrouper une ou plusieurs sous-classes de caractéristiques. Ces sous-classes sont le regroupement sémantique de quelques caractéristiques. Nous illustrons la lecture de notre modèle sur deux exemples :
- par rapport à l'agrégation possible des caractéristiques au niveau de la classe des caractéristiques relationnelles et contextuelles. Cette classe contient la sous-classe confiance, cette dernière regroupant quatre caractéristiques : la confiance envers le fournisseur, la confiance envers le client, la confiance envers les prestataires de services et l'engagement relationnel.
- Par rapport aux interactions entre caractéristiques. La caractéristique « Technologie du commerce électronique » (facteur), de la classe « caractéristiques technique et choix technologique », agit positivement sur la caractéristique « collaboration intra-organisationnelle » (impact/résultat), de la sous-classe collaboration. Cette caractéristique agit à son tour positivement (elle devient facteur) sur la caractéristique « performance organisationnelle » (impact/résultat) de la sous-classe performance de l'entreprise de la classe « caractéristiques de performance ».

Nous pouvons représenter le modèle sous une forme agrégée (Figure 2) selon le nombre de flèches reliant une classe à une autre. Ce modèle global regroupe évidemment les quatre classes de caractéristiques. Les interactions entre ces différentes classes sont présentées par des flèches pleines, l'épaisseur de la flèche étant proportionnelle au nombre d'hypothèses testées et vérifiées entre ces classes. Cette agrégation fait ressortir l'importance des recherches menées ces dernières années sur les facteurs explicatifs de la performance. Comme première interprétation, on peut remarquer que la classe de caractéristiques de performance est la classe 'cible'.

La classe des caractéristiques relationnelles et contextuelles représente le « noyau » du modèle puisqu'elle interagit avec toutes les autres classes. De plus elle présente la plus grande influence sur la classe de performance. Donc pour une entreprise, il est intéressant de se concentrer sur ce type de caractéristique et de définir ses relations et ses stratégies avec ses partenaires.

Aussi importante, la classe des caractéristiques techniques et du choix technologique a une assez forte influence sur les caractéristiques de performance. Cette classe interagit aussi avec celle des caractéristiques relationnelles mais pas avec celle des pratiques de la chaîne logistique.

Enfin, la classe des caractéristiques de pratiques de la chaîne a une influence moyenne sur la classe de performance et n'a aucune relation avec les caractéristiques techniques et choix technologiques, par contre elle parait être plutôt influencée par la classe des caractéristiques relationnelles et contextuelles.

L'apport de ces deux modèles est double. D'une part pour les chercheurs, ces modèles agrègent différents points de la performance (plusieurs types de caractéristiques : techniques, relationnelles, …), ce qui est quasiment absent de la littérature scientifique puisque on trouve généralement un ou deux types de caractéristiques étudiés dans chaque travail de recherche. En d'autres termes, ils proposent une vue générale des facteurs contribuant à la performance.

D'autre part, ces modèles permettent à une entreprise de se positionner par rapport à ses pratiques collaboratives et aux relations qu'elle entretient avec ses partenaires de la chaîne logistique. En fonction de ses objectifs de performance, elle pourra redéfinir ses relations avec ses partenaires, connaître l'apport et le gain dans l'investissement de technologies d'information et de communication, faire des choix technologiques adéquats à ses besoins et connaître la limite de quelques pratiques dans l'amélioration de la performance de la chaîne logistique.

Cependant, il ressort de nos analyses qu'aucune hypothèse testée et non vérifiée par un auteur n'a été étudiée par un autre. De plus notre modèle fait apparaître de nombreuses hypothèses qui n'ont pas encore été testées.

## 5 CONCLUSION

Pour rester compétitive, toute entreprise cherche à construire une stratégie lui permettant de se placer dans les meilleures conditions face aux forces concurrentielles présentes au sein de son secteur d'activité. En faisant l'hypothèse que la performance d'une entreprise, tant en termes de coût, de qualité et de délais, dépend de plus en plus fortement de sa capacité à optimiser ses relations avec ses partenaires et de ses dispositifs de collaboration et d'échange d'informations, ce travail a consisté à proposer un modèle des facteurs influençant la performance des différents acteurs dans les chaînes logistiques.

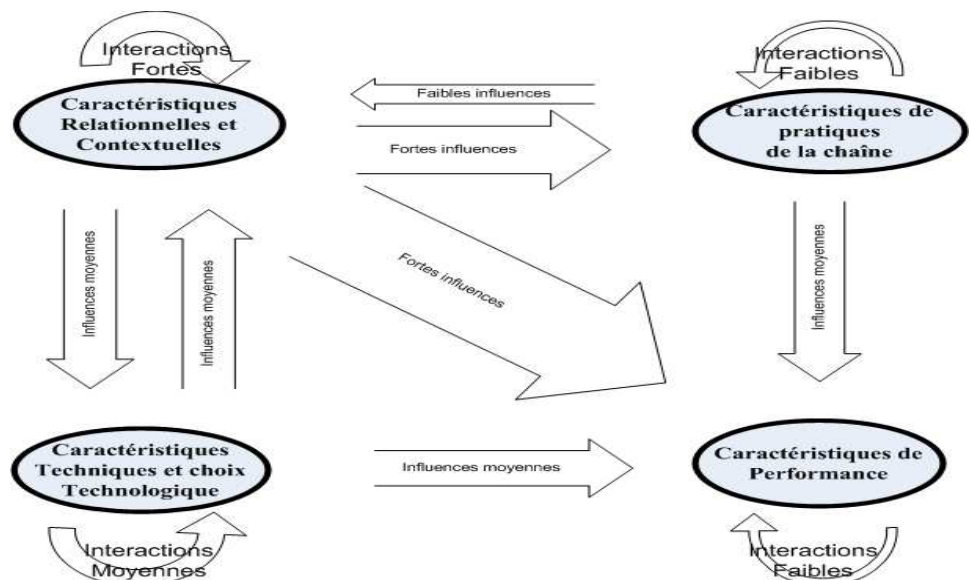

**Figure 2. Modèle général**

Nous avons identifié ainsi les principaux facteurs qui peuvent contribuer à la performance. Cette performance concerne soit chaque acteur de la chaîne en tant que tel, soit toute la chaîne logistique. Nous distinguons la performance de l'entreprise, du fournisseur, du client, ainsi que des éléments indispensables pour atteindre ces performances tels que l'usage des systèmes d'informations et autres technologies de communication, la collaboration intra et inter organisationnelle, la communication inter-organisationnelle, les pratiques de production,... Le modèle que nous proposons a été réalisé à partir d'une dizaine de travaux majeurs.

Chacun de ces auteurs s'est intéressé à un aspect de performance différent des autres. Notre modèle, qui regroupe 55 hypothèses sur 46 caractéristiques, représente un référentiel de facteurs contribuant à la performance des chaînes logistiques pour les chercheurs et pour les entreprises. Ce modèle pourrait bien évidemment être complété et enrichi par une analyse plus vaste de la littérature. En effet on peut s'interroger sur le rôle d'autres pratiques comme le VMI et le CPFR sur la performance.

En termes de perspectives, il serait intéressant de tester de nouvelles hypothèses et d'étudier la correspondance entre la configuration d'une entreprise (comme sa stratégie globale, les caractéristiques de ses lignes de produits) et les pratiques collaboratives qu'elle met ou souhaite mettre en œuvre au niveau de certains processus clés pour assurer la coordination entre les acteurs nécessaire à l'amélioration de leur performance.

Il serait également intéressant de tester la transitivité des relations entre les différentes caractéristiques, par exemple si « A *a un effet positif sur* B » et « B *a un effet positif sur* C », est ce que « A *a un effet positif sur* C » ? Ceci pourrait donner plus de choix aux preneurs de décision et éventuellement réduire le nombre de caractéristiques.

Enfin, la notion d'hypothèses pourrait être étendue aux hypothèses conditionnelles. Certaines hypothèses simples non vérifiées pourraient peut-être l'être sous condition : si « A n'*a pas d'effet positif sur* B » peut-on ajouter un facteur C tel que « A sous la condition C *a un effet positif sur* B » ?